\documentclass[aps,prb,groupedaddress,twocolumn,letterpaper,final,amssymb]{revtex4-2}

\usepackage{epsfig}
\usepackage{epstopdf}
\usepackage{color}
\usepackage{multirow}
\usepackage{dcolumn}
\begin{document}

\renewcommand{\dbltopfraction}{1.0}

\title{Stability, efficiency, and mechanism of $n$-type doping by hydrogen adatoms\\
in two-dimensional transition metal dichalcogenides}
\author{Sehoon Oh}
\author{June Yeong Lim}
\author{Seongil Im}
\author{Hyoung Joon Choi}
\email[Email:\ ]{h.j.choi@yonsei.ac.kr}
\affiliation{Department of Physics, Yonsei University, Seoul 03722, Korea}
\date{\today}

\begin{abstract}
Mono- and few-layer transition-metal dichalcogenides (TMDCs) provide opportunities for ideal two-dimensional semiconductors for electronic and optoelectronic devices. For electronic devices on TMDCs, it is essential to incorporate $n$- and/or $p$-type dopants which are stable in positions after patterned doping. Here we investigate hydrogen doping for TMDC ($MX_2$ with $M$ = Mo, W and $X$ = S, Se, Te) nanosheets by first-principles calculations to address diffusion and doping properties. We find that adsorbed hydrogen atoms in TMDCs are energetically most stable at the interstitial site right on the Mo or W plane and have substantial energy barriers against diffusion that increase in the order of sulfides, selenides, and tellurides. Located at the most stable interstitial site on the Mo or W plane, the hydrogen atoms produce electrons in the conduction bands in the extremely high rate of one electron per hydrogen atom, without any defect state inside the band gap remarkably. We analyze the chemical bonding character around the dopant and the mechanism for such high efficiency of electron doping. We also consider properties of hydrogen molecules and Te vacancies for comparison. Our work shows that hydrogen doping is the promising pathway to development of highly integrated electronic devices on TMDCs.
\end{abstract}

\maketitle

\section{Introduction}
\vspace{-2mm}
After the success of graphene, other two-dimensional materials such as transition-metal dichalcogenides (TMDCs) and black phosphorus have attracted enormous attention due to their potential in applications \cite{ref1,ref2,ref3,ref4,ref5,ref6,ref7,ref8,ref9,ref10,ref11,ref12,ref13,ref14,ref15,ref16,ref17,ref18,ref19,ref20,ref21,ref22,ref23,ref24,ref25,ref26,ref27,ref28}. Especially, TMDCs have been extensively investigated because of their sizable and tunable band gaps essential for switching devices \cite{ref1,ref2,ref3,ref4,ref5}, catalytic performance for hydrogen evolution reaction \cite{ref16,ref22}, and potentials in piezoelectric \cite{ref23,ref24}, optoelectronic \cite{ref25,ref26}, and spintronic devices \cite{ref27}.

To develop highly integrated devices based on TMDCs, it is essential to control the type and concentration of charge carriers locally in a patterned way. However, each TMDC typically showed a type of doping as a whole \cite{ref11,ref12,ref17}. Various point defects have been studied for TMDCs \cite{ref6,ref7,ref8,ref9,ref10,ref13,ref14,ref15}, including vacancies \cite{ref15} and adsorption of atoms such as potassium \cite{ref7}, rhenium \cite{ref8}, gold \cite{ref8}, niobium \cite{ref9}, hydrogen \cite{ref14,ref15,ref16,ref17,ref28}, chlorine \cite{ref17}, and second-row elements in the periodic table \cite{ref14,ref15}. Among these, the hydrogen adatom is a good candidate for the patterned doping on TMDCs \cite{ref28}.

The hydrogen doping in silicon and conventional bulk semiconductors has been studied intensively for its important role of passivating the electrical activity of dangling and defective bonds \cite{ref29,ref30,ref31,ref32}. Hydrogen impurities are often found to have the amphoteric behavior which counteracts the prevailing conductivity of the hosting semiconductor \cite{ref30,ref33}, but in some materials, hydrogen atoms act as a source of doping \cite{ref30,ref33}. For example, hydrogen atoms in bulk InN and ZnO are electron donors with defect levels above the conduction band minimum (CBM), while those in Ge, GaSb, and InSb are electron acceptors with defect levels below the valence band maximum (VBM) \cite{ref30,ref33}. For TMDCs, however, doped hydrogen atoms were often studied for their magnetic properties \cite{ref14,ref15}, while their charge doping effects were revealed only superficially \cite{ref14,ref17,ref28}.

The diffusivity of the dopant is one of the key concerns that affect the design, fabrication, and applicability of the devices. For its importance, hydrogen diffusion has been widely studied for conventional bulk semiconductors \cite{ref31,ref32,ref34}. In the case of TMDCs, however, hydrogen diffusion has not been studied yet to the best of our knowledge. For the hydrogen atom to be useful as a dopant in TMDCs, it should have an energetically stable position with a substantial energy barrier that suppresses diffusion.

In the present work, we investigate doped hydrogen atoms in TMDC nanosheets by first-principles calculations based on the density functional theory (DFT). Our calculations show that the doped hydrogen atom is most stable at the center of the interstitial site right on the transition-metal plane, and that the diffusion barrier is high enough in selenides and tellurides to suppress the spatial diffusion of the hydrogen atom. Furthermore, the hydrogen atoms at our lowest-energy configuration do not make any defect state inside the band gap, and produce electrons in the conductions bands extremely efficiently, in the rate of one electron per hydrogen atom, unlike the hydrogen atoms adsorbed to chalcogen atoms. We analyze the chemical bonding character of the hydrogen atom at the interstitial site and present the mechanism for its highly efficient $n$-type doping. Hydrogen molecules and Te vacancies are also considered for doping effects, for comparison. Our results show that hydrogen atoms are excellent for position-selective $n$-type doping to TMDCs, opening a chance for stably patterned electron doping in a single TMDC nanosheet.

\begin{table*}[t] 
\caption
[Structural and energetic properties of $MX_2$ monolayer.]
{Structural and energetic properties of $MX_2$ monolayer. The in-plane lattice constant, $a$, and the interlayer distance, $d_{M\text{-}X}$, between the transition-metal layer and the chalcogen layer are obtained theoretically for pristine 1H-phase MoS$_2$, MoSe$_2$, MoTe$_2$, WS$_2$, WSe$_2$, and WTe$_2$ monolayers. The binding energies, $E_\mathrm{bind}$, are the amount of energy released per hydrogen atom when it is adsorbed at the hollow, $M$-top, and $X$-top sites, respectively, from monoatomic gas phase. The distance, $d_{H\text{-}M}$ ($d_{H\text{-}X}$), is from the hydrogen atom to the nearest transition-metal (chalcogen) atom in the $M$-top ($X$-top) case. The diffusion barrier height, $E_\mathrm{barr}$, of the hydrogen atom from the lowest-energy adsorption site is $E_\mathrm{bind}$[hollow]$-E_\mathrm{bind}$[$X$-top] for MoS$_2$ and WS$_2$ monolayers, and $E_\mathrm{bind}$[hollow]$-E_\mathrm{bind}$[$M$-top] for MoSe$_2$, MoTe$_2$, WSe$_2$, and WTe$_2$ monolayers. Here, the binding energies are obtained using a $5\times5$ supercell with one H atom in it.}
\label{tab1}
\setlength{\tabcolsep}{7.3mm} 
\renewcommand{\arraystretch}{1.12}
\begin{tabular}{l c c c c c c}
\hline\hline
& MoS$_2$ & MoSe$_2$ & MoTe$_2$ & WS$_2$ & WSe$_2$ & WTe$_2$ \\
\hline
$a$ ({\AA}) & 3.279 & 3.391 & 3.624 & 3.279 & 3.411 & 3.634 \\
$d_{M\text{-}X}$ ({\AA}) & 1.619 & 1.711 & 1.854 & 1.613 & 1.702 & 1.843 \\
$E_\mathrm{bind}$[hollow](eV) & 1.700 & 2.066 & 2.614 & 1.635 & 2.079 & 2.686 \\
$E_\mathrm{bind}$[$M$-top](eV) & 1.266 & 1.400 & 1.769 & 1.056 & 1.342 & 1.760 \\
$E_\mathrm{bind}$[$X$-top](eV) & 1.557 & 1.338 & 1.004 & 1.282 & 1.132 & 1.199 \\
$d_{H\text{-}M}$ ({\AA}) & 1.762 & 1.760 & 1.754 & 1.752 & 1.755 & 1.759 \\
$d_{H\text{-}X}$ ({\AA}) & 1.424 & 1.582 & 1.731 & 1.450 & 1.621 & 1.732 \\
$E_\mathrm{barr}$ (eV) & 0.143 & 0.666 & 0.846 & 0.353 & 0.737 & 0.926 \\
\hline\hline
\end{tabular}
\end{table*} 

\vspace{-2mm}
\section{Computational Methods}
\vspace{-2mm}
We investigate the atomic and electronic structures of 1H-phase $MX_2$ ($M$ = Mo, W; $X$ = S, Se, Te) monolayers by performing first-principles calculations based on the DFT. We use the generalized gradient approximation \cite{ref35}, norm-conserving pseudopotentials \cite{ref36}, and localized pseudoatomic orbitals for wave functions, as implemented in the SIESTA code \cite{ref37}. Spin-orbit interaction, which is considered in structure optimizations and electronic structure calculations, is incorporated within fully relativistic $j$-dependent pseudopotentials \cite{ref38} in the $l$-dependent fully separable nonlocal form using additional Kleinman-Bylander-type projectors \cite{ref39,ref40}. Our method was successfully used for the topological insulators \cite{ref41} and for solids and nanostructures with giant spin splitting \cite{ref42}. We use $32\times32\times1$ Monkhorst-Pack $k$-point mesh for all supercell calculations of monolayer and bilayer, and $16\times16\times16$ $k$-point mesh for bulk $MX_2$. Real-space mesh cutoff of 2000 Ry is used for all of our calculations. DFT-D2 correction is applied to account for the van der Waals interaction. The binding energy of a hydrogen atom or molecule is obtained by
\begin{eqnarray} 
 E_\mathrm{bind} &=& E_\mathrm{tot}\text{(pristine $MX_2$)} + E_\mathrm{tot}\text{(isolated H or H$_2$)}\nonumber\\
&& - E_\mathrm{tot}\text{(H-~or~H$_2$-adsorbed $MX_2$)},\nonumber
\end{eqnarray}
where terms on the right-hand side are the total energies of pristine $MX_2$, isolated H or H$_2$, and H- or H$_2$-adsorbed $MX_2$, respectively. We also use the Vienna ab initio simulation package (VASP) code \cite{ref43,ref44} to double-check our SIESTA results \cite{ref45} and to confirm the previously reported calculation results in the literature \cite{ref14,ref16}. Since the SIESTA code uses local orbitals, we estimated typical size of the basis set superposition error (BSSE) in the binding energy by introducing ghost atoms in our SIESTA calculations, and found that BSSE typically increases or decreases the binding energy by about 0.01~eV, which is small enough not to change any conclusion in our present work.

\begin{figure*}[t] 
\includegraphics[width=15.0cm]{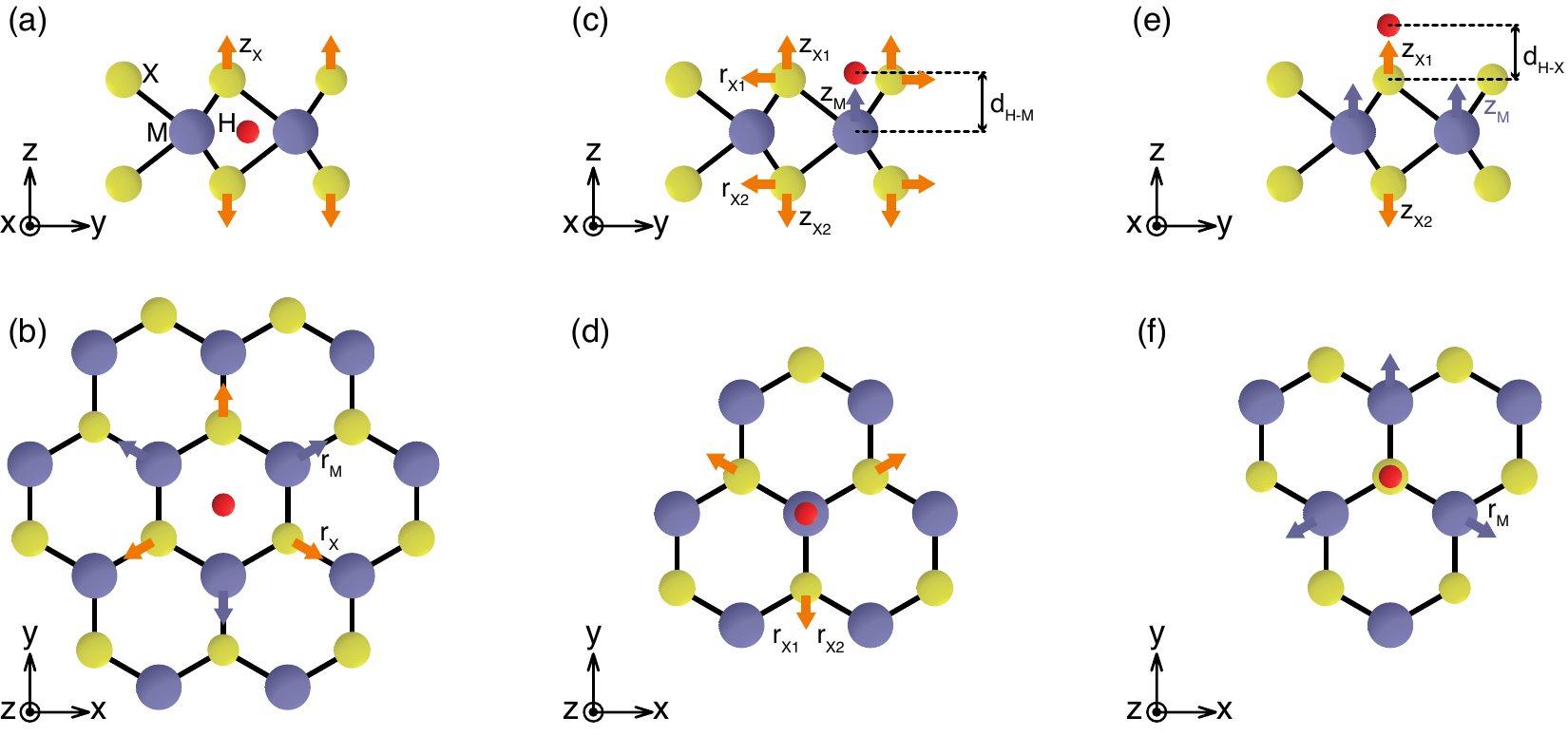}
\caption
[Atomic displacements in TMDC ($MX_2$) monolayer induced by an adsorbed hydrogen atom]
{Atomic displacements in TMDC ($MX_2$) monolayer induced by an adsorbed hydrogen atom at (a), (b) the center of the hexagonal hollow site (hollow), (c), (d) the top of the transition-metal atom ($M$-top), and (e), (f) the top of the chalcogen atom ($X$-top). Top (a), (c), (e) and side views (b), (d), (f) are shown. Blue, yellow, and red spheres represent transition-metal, chalcogen, and hydrogen atoms, respectively. The hydrogen atom is located right on the transition-metal plane in the hollow case. Blue and orange arrows show displacements of transition-metal and chalcogen atoms induced by the hydrogen-atom adsorption, respectively. In (a), $z_X$ is the out-of-plane displacement of the nearest chalcogen atoms from the hydrogen atom. In (b), $r_M$ and $r_X$ are the radial displacement of the nearest transition-metal and chalcogen atoms, respectively. In (c), $z_M$, $z_{X1}$, and $z_{X2}$ are out-of-plane displacements of the nearest transition-metal atom, the nearest chalcogen atoms, and the second nearest chalcogen atoms, respectively. In (c) and (d), $r_{X1}$ and $r_{X2}$ are radial displacements of the nearest and the second nearest chalcogen atoms, respectively. In (e), $z_M$, $z_{X1}$, and $z_{X2}$ are out-of-plane displacements of the nearest transition-metal atoms and the upper and lower chalcogen atoms, respectively. In (f), $r_M$ is the radial displacement of the nearest transition-metal atoms. Numerical values of these atomic displacements are shown in Table~\ref{tab2}.}
\label{fig1}
\end{figure*}

The lattice parameters of pristine $MX_2$ were determined by minimizing the total energy. The obtained structural parameters for monolayers are summarized in Table~\ref{tab1}. Then we construct a $3\times3$ supercell of $MX_2$ monolayer with an adsorbed hydrogen atom in it and optimized all atomic positions in the supercell by minimizing the total energy. During this atomic structure relaxation, the lattice constants are fixed to the values of the pristine case, and the vacuum region in the supercell is $\sim$ 100-{\AA} thick along the direction perpendicular to the $MX_2$ layer in order to avoid fictitious interactions between layers generated by the periodic boundary condition. The atomic positions are fully optimized until residual forces are less than 0.005~eV/\AA. The spin-orbit and van der Waals interactions are considered during the atomic structure optimization. During this relaxation, only atoms near the hydrogen atom are shifted significantly (Table~\ref{tab2} and Fig.~\ref{fig1}), while the other atoms are hardly shifted from their original positions. Then in order to reduce the interaction between a hydrogen atom and its periodic images in the supercell calculation we construct a $5\times5$ supercell of $MX_2$ monolayer with an adsorbed hydrogen atom in it by replacing positions of the hydrogen atom and its nearby atoms with relaxed atomic positions from the $3\times3$ supercell calculation. With this geometry of the $5\times5$ supercell we calculate the total energy, the electronic band structure, the projected density of states, the wave functions, and the charge density.

\begin{table*}[t] 
\caption
[Atomic displacements in $MX_2$ monolayer induced by an adsorbed hydrogen atom at hollow, $M$-top, and $X$-top sites.]
{Atomic displacements in $MX_2$ monolayer induced by an adsorbed hydrogen atom at hollow, $M$-top, and $X$-top sites. The displacements $r_M$, $r_X$ , $z_X$ , $z_M$, $r_{X1}$, $r_{X2}$, $z_{X1}$, and $z_{X2}$ are defined in Fig.~\ref{fig1}.}
\label{tab2}
\setlength{\tabcolsep}{6.1mm} 
\renewcommand{\arraystretch}{1.12}
\begin{tabular}{l c r r r r r r}
\hline\hline
& & MoS$_2$ & MoSe$_2$\hspace{-1mm} & MoTe$_2$\hspace{-1mm} & WS$_2$\hspace{0.5mm} & WSe$_2$ & WTe$_2$\hspace{-0.5mm} \\
\hline
\multirow{3}*{Hollow} & $r_M$(\AA) & 0.042 & 0.027 & $-$0.034 & 0.066 & 0.041 & $-$0.008 \\
& $r_X$ (\AA) & 0.031 & 0.044 & 0.034 & 0.026 & 0.044 & 0.030 \\
& $z_X$ (\AA) & 0.013 & 0.025 & 0.033 & 0.006 & 0.004 & 0.023 \\
\hline
\multirow{5}*{$M$-top} & $z_M$ (\AA) & 0.066 & 0.068 & 0.095 & 0.103 & 0.080 & 0.096 \\
& $r_{X1}$ (\AA) & 0.102 & 0.105 & 0.134 & 0.130 & 0.127 & 0.136 \\
& $r_{X2}$ (\AA) & $-$0.008 & $-$0.011 & $-$0.011 & $-$0.015 & $-$0.011 & $-$0.012 \\
& $z_{X1}$ (\AA) & 0.029 & 0.035 & 0.017 & 0.023 & 0.005 & 0.014 \\
& $z_{X2}$ (\AA) & $-$0.037 & $-$0.032 & $-$0.049 & $-$0.056 & $-$0.040 & $-$0.052 \\
\hline
\multirow{4}*{$X$-top} & $z_{X1}$ (\AA) & $-$0.011 & 0.011 & $-$0.381 & $-$0.001 & $-$0.032 & $-$0.274 \\
& $z_{X2}$ (\AA) & 0.015 & 0.004 & $-$0.133 & 0.037 & 0.006 & $-$0.091 \\
& $r_M$ (\AA) & $-$0.005 & $-$0.008 & 0.220 & $-$0.022 & 0.009 & 0.122 \\
& $z_{M}$ (\AA) & 0.007 & 0.022 & $-$0.005 & 0.005 & $-$0.003 & $-$0.006 \\
\hline\hline
\end{tabular}
\end{table*} 

\begin{figure} 
\includegraphics[width=7.0cm]{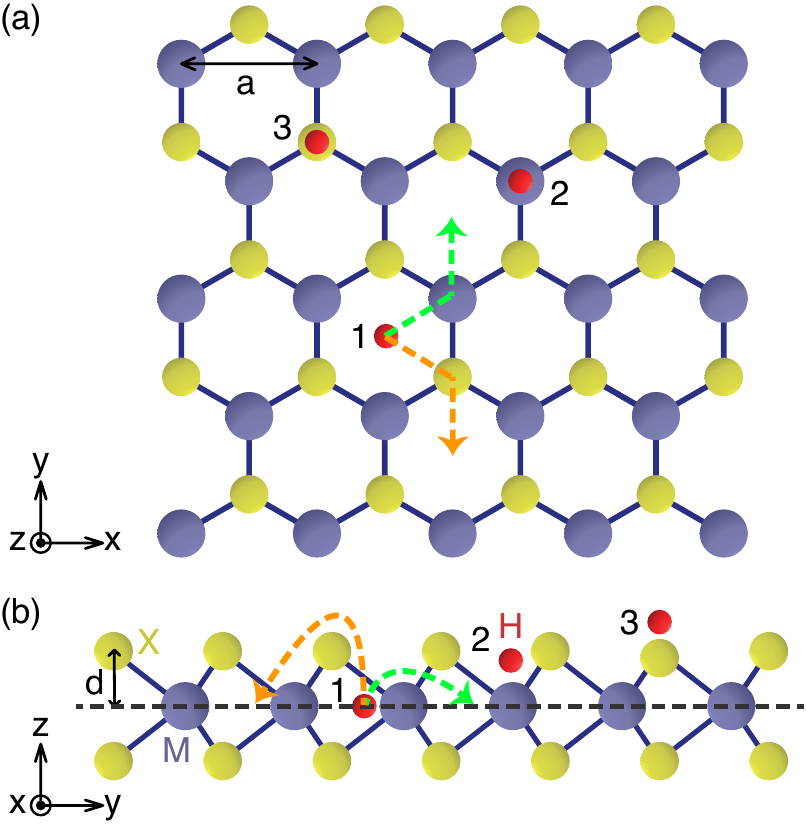}
\caption
[Schematic diagram for atomic structure and diffusion paths in $MX_2$ monolayer with adsorbed hydrogen atoms.]
{Schematic diagram for atomic structure and diffusion paths in $MX_2$ monolayer with adsorbed hydrogen atoms. (a) Top view and (b) side view. Blue, yellow, and red spheres represent transition- metal, chalcogen, and hydrogen atoms, respectively. The adsorbed hydrogen atoms at the hollow, $M$-top, and $X$-top sites are denoted by numbers in order. The hollow site is the lowest-energy adsorption site. The lowest-energy path from a hollow site to its neighboring site is through the $M$-top site in the cases of MoSe$_2$, MoTe$_2$, WSe$_2$, and WTe$_2$ monolayers, as represented by green dashed arrows, and it is through the $X$-top site in the cases of MoS$_2$ and WS$_2$ monolayers, as represented by orange dashed arrows.}
\label{fig2}
\end{figure} 

\begin{table}[t] 
\caption
[Structural and energetic properties of $MX_2$ bulk.]
{Structural and energetic properties of $MX_2$ bulk. The in-plane lattice constant, $a$, the out-of-plane lattice constant, $c$, and the interlayer distance, $d_{M\text{-}X}$ , between the transition-metal layer and the chalcogen layer are obtained theoretically for pristine 2H-phase MoS$_2$, MoSe$_2$, MoTe$_2$, WS$_2$, WSe$_2$, and WTe$_2$ bulk. The binding energies, $E_\mathrm{bind}$, are the amount of energy released per hydrogen atom when it is adsorbed at the hollow, $M$-top, and $X$-top sites, respectively, from monoatomic gas phase. The distance, $d_{H\text{-}M}$ ($d_{H\text{-}X}$), is from the hydrogen atom to the nearest transition-metal (chalcogen) atom in the $M$-top ($X$-top) case. The lowest energy adsorption site is the $X$-top site in MoS$_2$ bulk and the hollow site in the other bulk compounds. The diffusion barrier height, $E_\mathrm{barr}$, of the hydrogen atom from the lowest-energy adsorption site is $E_\mathrm{bind}$[$X$-top]$-E_\mathrm{bind}$[hollow] in MoS$_2$ bulk, $E_\mathrm{bind}$[hollow]$-E_\mathrm{bind}$[$X$-top] in MoSe$_2$ and WS$_2$ bulk, and $E_\mathrm{bind}$[hollow]$-E_\mathrm{bind}$[$M$-top] in MoTe$_2$, WSe$_2$, and WTe$_2$ bulk. Here, the binding energies are obtained using a $3\times3$ supercell with one H atom in it.}
\label{tab3}
\setlength{\tabcolsep}{0.5mm} 
\renewcommand{\arraystretch}{1.12}
\begin{tabular}{l r r r r r r}
\hline\hline
& MoS$_2$ & MoSe$_2$\hspace{-1mm} & MoTe$_2$\hspace{-1mm} & WS$_2$\hspace{0.5mm} & WSe$_2$ & WTe$_2$\hspace{-0.5mm} \\
\hline
$a$ ({\AA}) & 3.278 & 3.389 & 3.623 & 3.278 & 3.409 & 3.630 \\
$c$ ({\AA}) & 12.499 & 13.112 & 14.102 & 12.189 & 12.947 & 13.919 \\
$d_{M\text{-}X}$ ({\AA}) & 1.618 & 1.712 & 1.852 & 1.608 & 1.695 & 1.840 \\
$E_\mathrm{bind}$[hollow](eV) & 1.779 & 2.206 & 2.730 & 1.716 & 2.167 & 2.722 \\
$E_\mathrm{bind}$[$M$-top](eV) & 1.371 & 1.570 & 1.883 & 1.183 & 1.469 & 1.830 \\
$E_\mathrm{bind}$[$X$-top](eV) & 1.834 & 1.619 & 1.599 & 1.537 & 1.354 & 1.323 \\
$d_{H\text{-}M}$ ({\AA}) & 1.783 & 1.785 & 1.772 & 1.772 & 1.767 & 1.768 \\
$d_{H\text{-}X}$ ({\AA}) & 1.421 & 1.569 & 1.739 & 1.417 & 1.581 & 1.754 \\
$E_\mathrm{barr}$ (eV) & 0.054 & 0.587 & 0.848 & 0.179 & 0.698 & 0.892 \\
\hline\hline
\end{tabular}
\end{table} 

\begin{figure} 
\includegraphics[width=7.0cm]{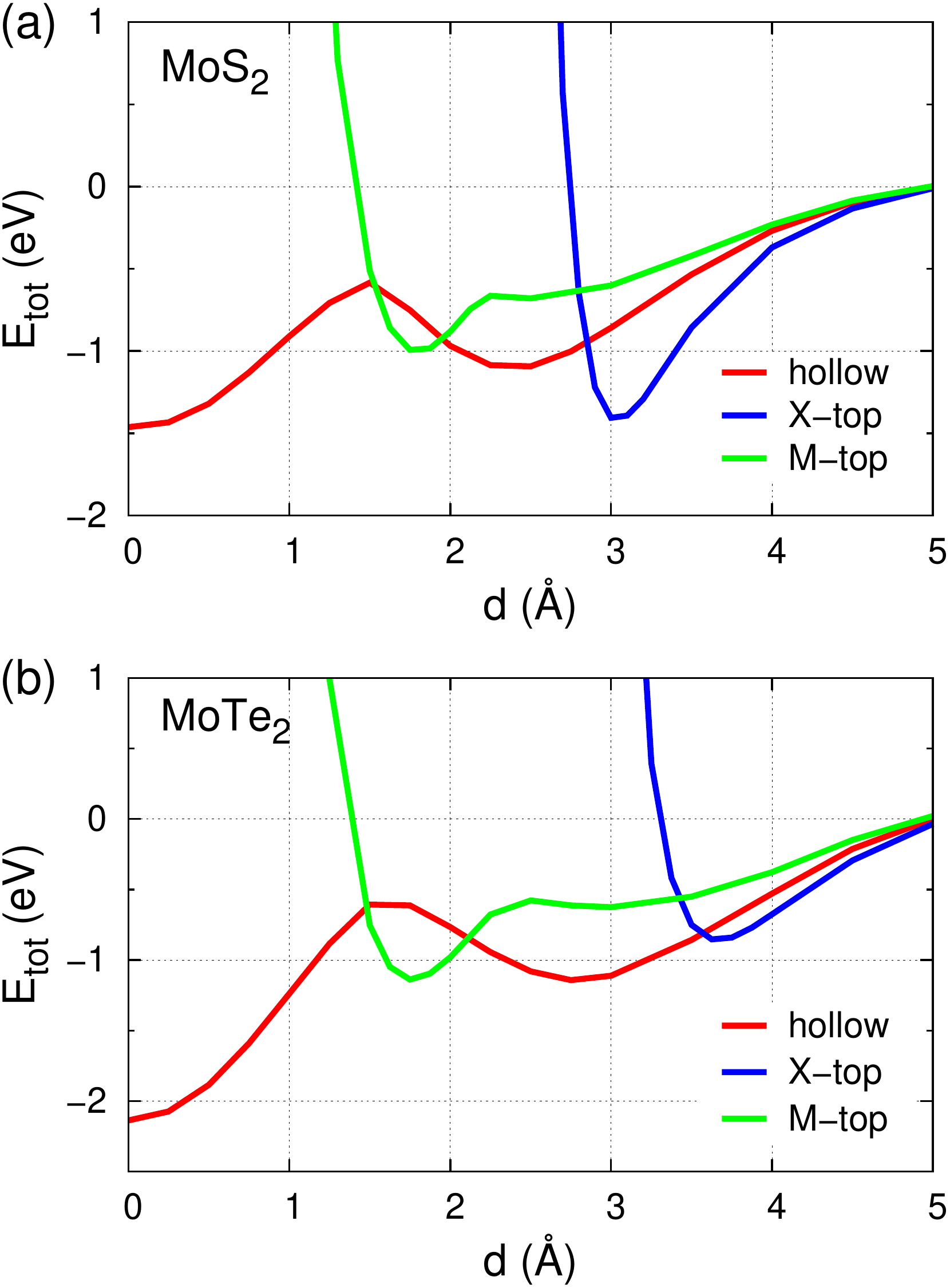}
\caption
[Total energy of (a) MoS$_2$ monolayer and (b) MoTe$_2$ monolayer with an adsorbed hydrogen.]
{Total energy of (a) MoS$_2$ monolayer and (b) MoTe$_2$ monolayer with an adsorbed hydrogen. The total energy of the system is plotted as a function of the vertical distance, $d$, of the hydrogen atom from the Mo plane. The total energy is set to zero for the hollow case of $d=5$~\AA. The red, blue, and green lines represent the total energy of the system as the hydrogen atom approaches the Mo plane vertically through the hollow, $X$-top, and $M$-top sites, respectively. For these plots, $5\times5$ supercells are used and Mo, S, and Te atoms are fixed at their positions as in pristine monolayers while the H atom is moved.}
\label{fig3}
\end{figure} 

\vspace{-2mm}
\section{Results and Discussions}
\vspace{-2mm}
\subsection{Binding energies and diffusion barriers of adsorbed hydrogen atom}
\vspace{-2mm}


We performed first-principles density-functional calculations of 1H-phase $MX_2$ ($M$ = Mo, W; $X$ = S, Se, Te) monolayers using a $5\times5$ supercell with one adsorbed hydrogen atom in it. As for the adsorption site of the hydrogen impurity, we considered a uniform grid in real space including high-symmetry sites and found that global or local minima in energy occur only at three high-symmetry sites, which are the center of the hexagonal hollow site (hollow), the top of the transition-metal atom ($M$-top), and the top of the chalcogen atom ($X$-top), as shown in Figs.~1 and 2. For all the investigated $MX_2$ monolayers, we found that the hollow site is the lowest-energy site and the binding energy of the hydrogen atom at the hollow site, which is the energy released in the adsorption process from the monoatomic gas phase, is $1.6\sim2.7$~eV as summarized in Table~\ref{tab1}. In our work, the lowest-energy hollow site is right on the transition-metal plane as shown in Fig.~\ref{fig2}(b), while the hollow site in Ref.~[15] is away from the metal plane. The hollow and $X$-top binding energies are quite comparable to each other in MoS$_2$ monolayer (Table~\ref{tab1}), but the hydrogen atoms are much more stable at the hollow site on the metal plane than at the $X$-top site in selenides and tellurides (Table~\ref{tab1}). Similar behaviors also occur in bulk case (Table~\ref{tab3}).

As shown in Fig.~\ref{fig3}, when a hydrogen atom approaches the hollow site of MoS$_2$ or MoTe$_2$ monolayer from vacuum, there exists a local minimum of the total energy at 2~$\sim$~3~{\AA} away from the Mo plane and this local minimum can be higher than or comparable to the $X$- or $M$-top energy. Thus, during structural optimization the true total-energy minimum right on the Mo plane can be missed if sampling of the initial hydrogen-atom position does not include a position close to the center of the hollow site right on the metal plane.

To investigate spatial stability of hydrogen atoms, we calculated the total energy of the system as a function of the hydrogen-atom position from one hollow site to one of its neighboring hollow sites. We found that the lowest-energy diffusion path is through either the $M$-top or the $X$-top site (Fig.~\ref{fig2}). We also tried different paths such as moving through between two chalcogen atoms or between a metal and a chalcogen atom, but these paths cost larger energy than those through the $M$- or $X$-top site. Since the $M$- and $X$-top sites are local minima in the total energy, the maximal barrier height in the diffusion path through the $M$- or $X$-top site is greater than the binding-energy difference between the hollow site and the $M$- or $X$-top site. In Table~\ref{tab1}, the barrier height is defined as the binding-energy difference between the hollow site and the $M$- or $X$-top site, which is $0.14\sim0.93$~eV depending on material compositions, so the maximal barrier height along the diffusion path is greater than these values. In MoSe$_2$, MoTe$_2$, WSe$_2$, and WTe$_2$ monolayers, the $M$-top site is lower in energy than the $X$-top site, while it is reversed in MoS$_2$ and WS$_2$ monolayers (Table~\ref{tab1}). Especially, the barrier height is $0.67\sim0.93$~eV in selenides and tellurides, which is much greater than the room temperature scale, making adsorbed hydrogen atoms hardly mobile at room temperature. Since the diffusion barrier as well as the binding energy increases in the order of sulfide, selenide, and telluride compounds, hydrogen atoms are most stable in telluride compounds.

We also calculated the vibrational frequencies of the adsorbed hydrogen atom at the hollow site in MoTe$_2$ monolayer. Because of the large mass difference between the molybdenum and hydrogen atoms, Mo atoms can be regarded stationary when the hydrogen atom oscillates in space. The calculated characteristic frequencies of the out-of-plane and in-plane modes are 21.7~THz (724~cm$^{-1}$) and 24.9~THz (831~cm$^{-1}$), respectively.

\begin{figure} 
\includegraphics[width=8.0cm]{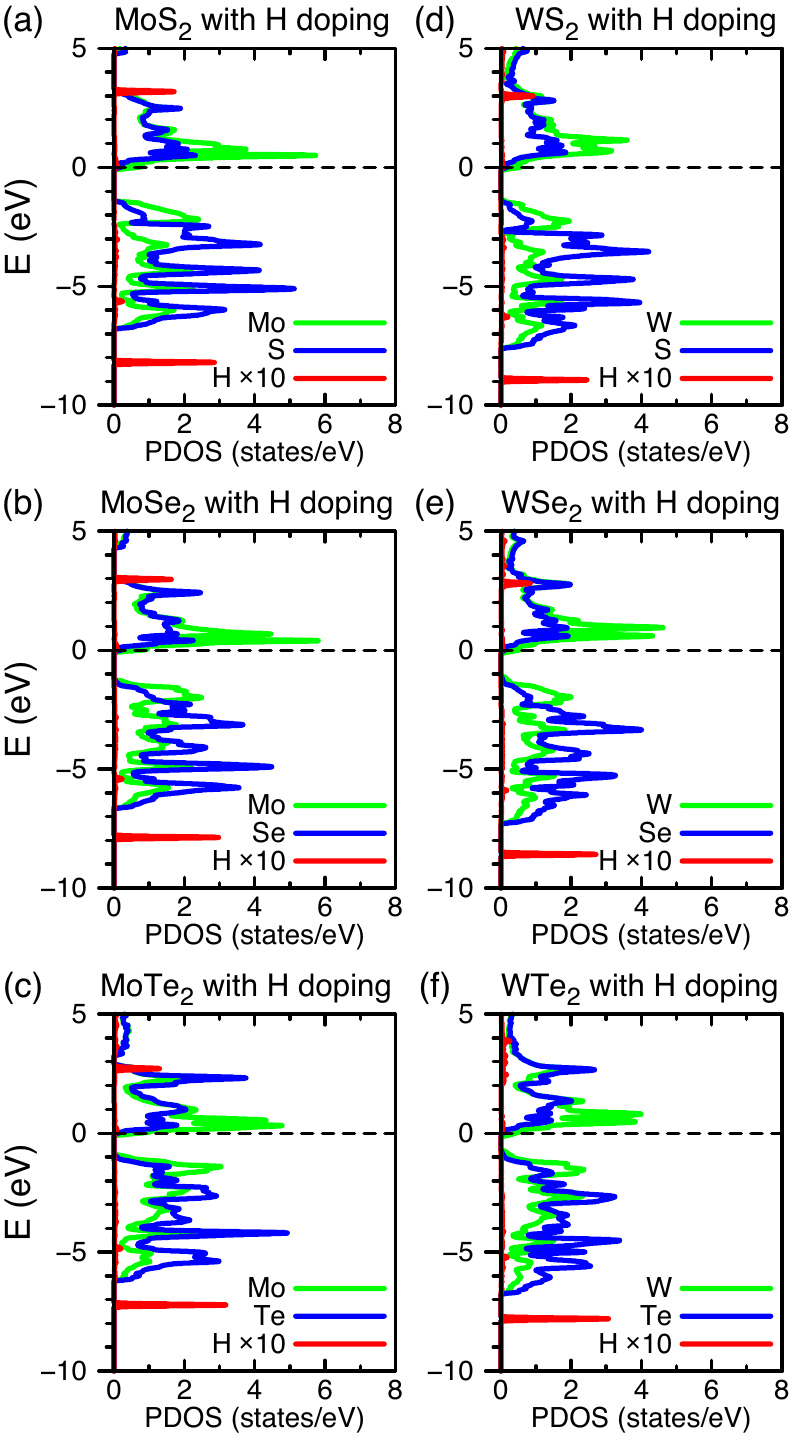}
\caption
[Electronic structures of $MX_2$ monolayers with adsorbed hydrogen atom.]
{Electronic structures of $MX_2$ monolayers with adsorbed hydrogen atom. (a)-(f), Projected density of states (PDOS) of MoS$_2$, MoSe$_2$, MoTe$_2$, WS$_2$, WSe$_2$, and WTe$_2$ monolayers with an adsorbed hydrogen atom at the hollow site. In each plot, the chemical potential is set to zero and marked with the horizontal dashed line. The hydrogen-atom concentration is 0.04 per $MX_2$ formula unit. The PDOS onto the hydrogen site is multiplied by a factor of 10, for clearer presentation.}
\label{fig4}
\end{figure} 

\begin{figure*}[t] 
\includegraphics[width=17.5cm]{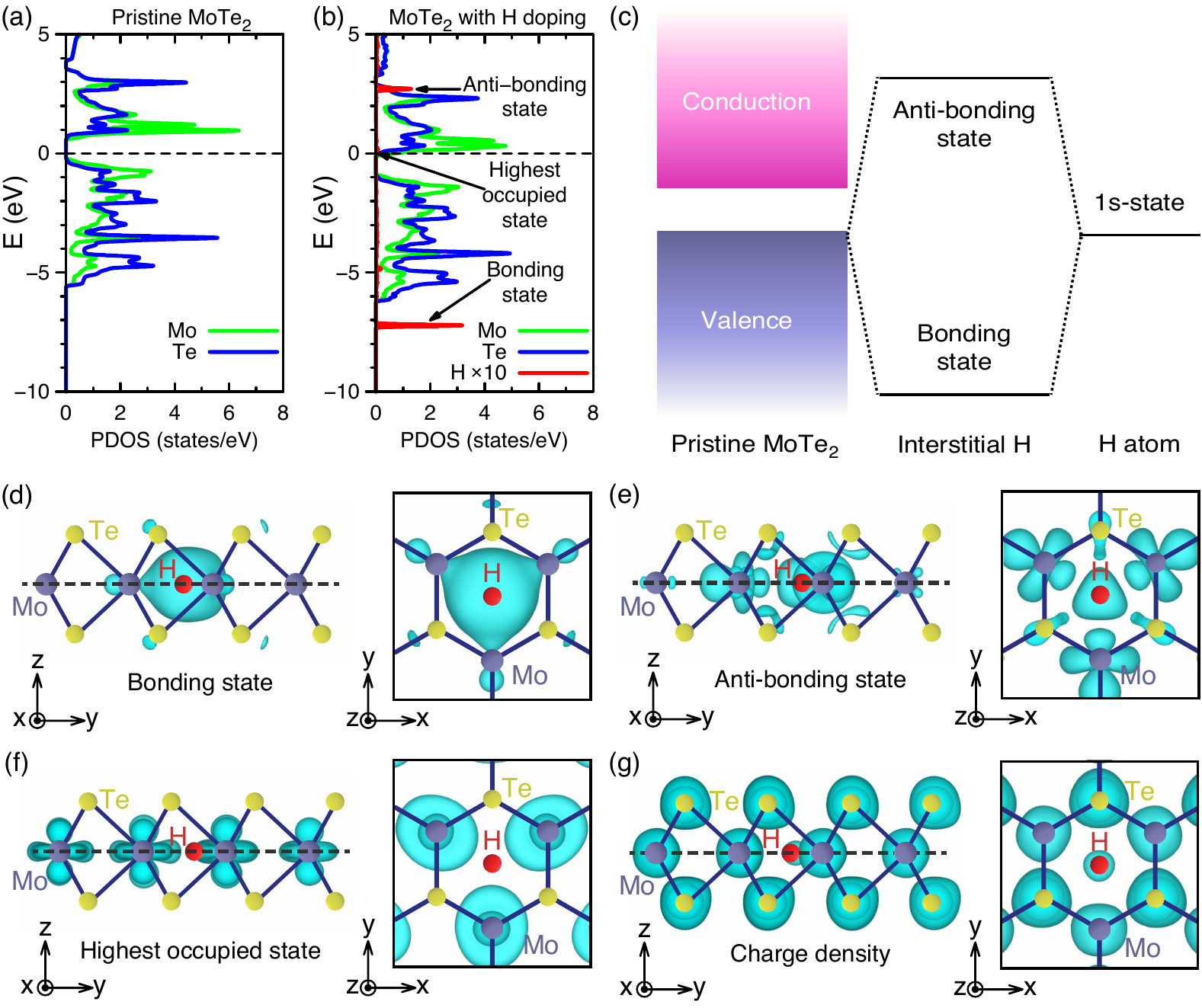}
\caption
[Doping mechanism in MoTe$_2$ monolayer with adsorbed hydrogen atom.]
{Doping mechanism in MoTe$_2$ monolayer with adsorbed hydrogen atom. (a) PDOS of pristine MoTe$_2$ monolayer. (b) PDOS of MoTe$_2$ monolayer with adsorbed hydrogen atom. In (a) and (b), the chemical potential is set to zero energy. (c) Schematic energy diagram for the formation of bonding and antibonding states. (d)-(f) Isosurface plots of the squared wave functions of the bonding state, antibonding state, and the highest occupied state in H-doped MoTe$_2$ monolayer, respectively. The highest occupied state corresponds to the CBM state of pristine MoTe$_2$. (g) Isosurface plot of the total charge density of H-doped MoTe$_2$.}
\label{fig5}
\end{figure*} 

\begin{figure} 
\includegraphics[width=8.0cm]{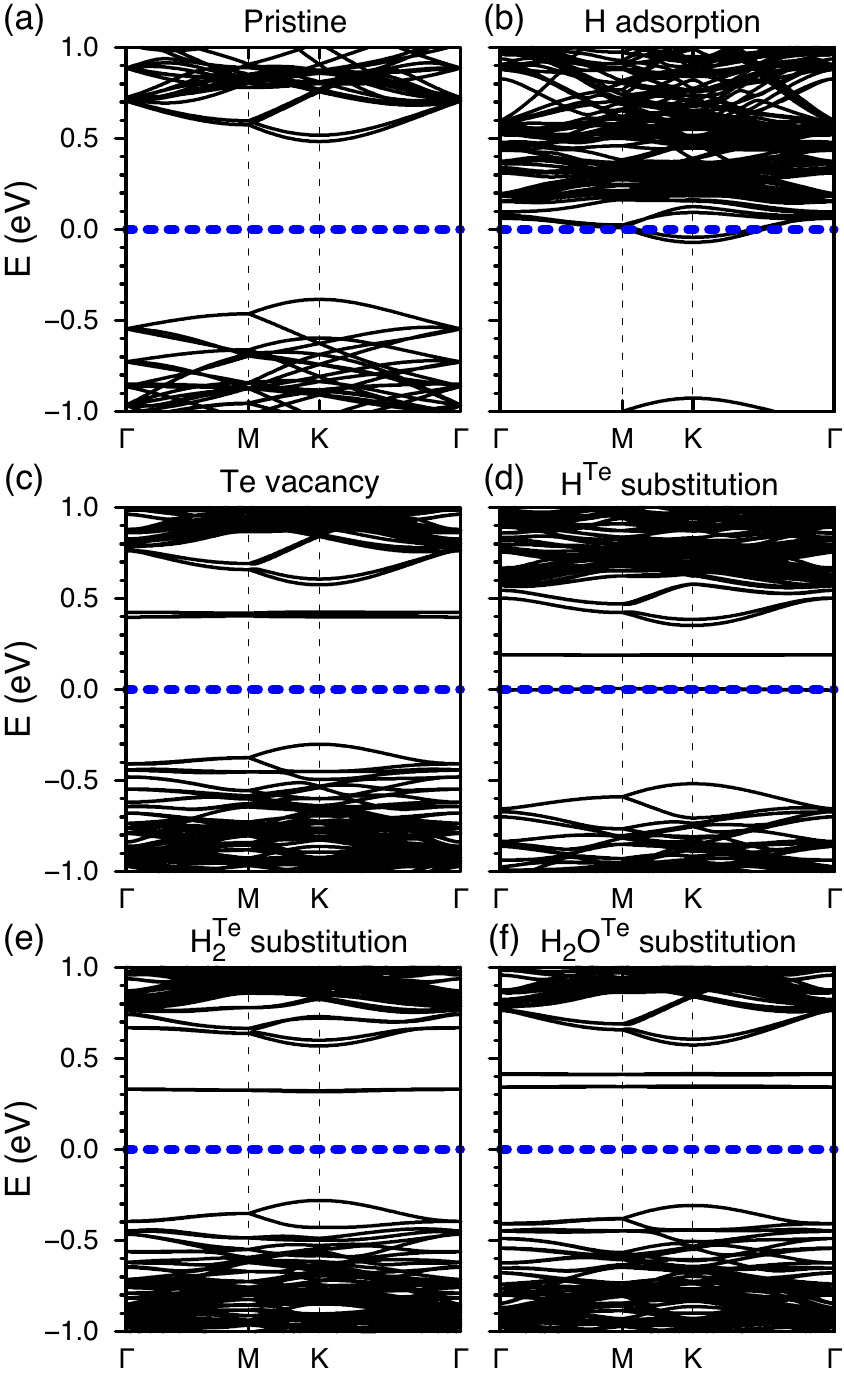}
\caption
[Band structures of MoTe$_2$ monolayer with defects.]
{Band structures of MoTe$_2$ monolayer with defects. (a) Pristine MoTe$_2$ monolayer drawn for comparison. (b) MoTe$_2$ monolayer with an interstitial hydrogen atom (H) at the hollow site. (c) MoTe$_2$ monolayer with a Te vacancy. (d)-(f) MoTe$_2$ monolayers with a Te atom replaced with a hydrogen atom (H), a hydrogen molecule (H$_2$), and a water molecule (H$_2$O), respectively. In each plot, the chemical potential is set to zero and marked with the blue dashed line. In (a)-(f), the Brillouin zone corresponds to a $5\times5$ supercell. In (b)-(f), the number of the defect is one in the $5\times5$ supercell.}
\label{fig6}
\end{figure} 

\begin{figure}[t] 
\includegraphics[width=6.5cm]{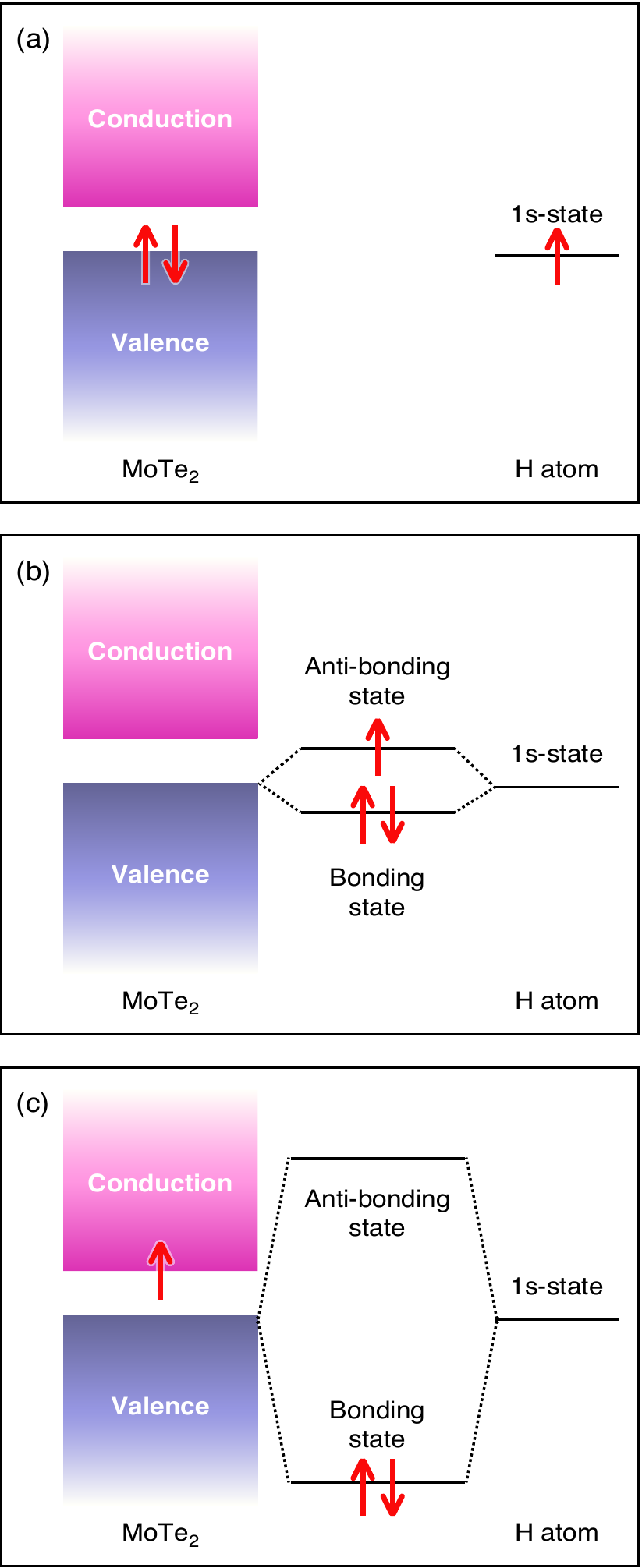}
\caption
[Schematic diagram of electron doping in H-doped MoTe$_2$.]
{Schematic diagram of electron doping in H-doped MoTe$_2$. (a) Before doping, electrons (shown as arrows) are in valence bands of MoTe$_2$ and the 1$s$ state of the H atom. (b) As the H atom approaches MoTe$_2$, bonding and antibonding states are formed. (c) When the H atom is at the hollow site on the Mo plane of MoTe$_2$, one electron moves from the antibonding state to conduction bands because the antibonding state is higher in energy than the conduction band minimum, resulting in one electron doping to conduction bands.}
\label{fig7}
\end{figure} 

\begin{figure*}[t] 
\includegraphics[width=17.0cm]{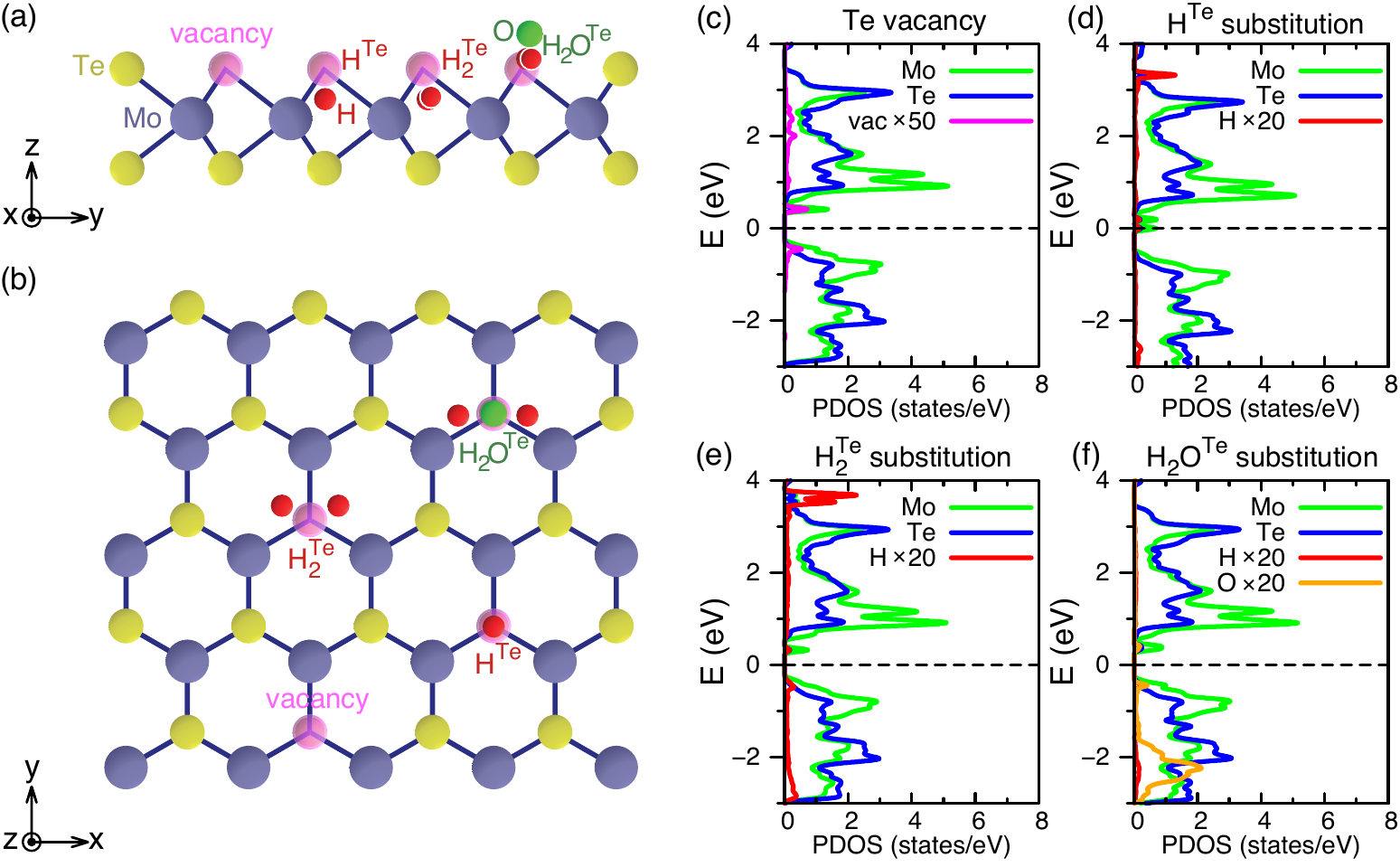}
\caption
[MoTe$_2$ monolayer with complex defects.]
{MoTe$_2$ monolayer with complex defects. (a), (b) Side and top views of the schematic atomic structures of MoTe$_2$ monolayer with complex defects. Blue, yellow, red, and green spheres represent Mo, Te, H, and O atoms, respectively, and magenta spheres represent Te vacancies. (c)-(f) PDOS of MoTe$_2$ monolayer with a Te vacancy, a H atom at the Te vacancy, a H$_2$ molecule at the Te vacancy, and a H$_2$O molecule at the Te vacancy, respectively. In each plot, the chemical potential is set to zero energy and marked with the dashed line.}
\label{fig8}
\end{figure*} 

\vspace{-2mm}
\subsection{Doping effects and doping mechanism}
\vspace{-2mm}
To find out doping effects, we analyzed the electronic density of states in $MX_2$ monolayers with a hydrogen atom adsorbed at the hollow site (Fig.~\ref{fig4}). Our results show that the hydrogen doping induces an electron-doping in $MX_2$ monolayers, raising the chemical potential somewhat higher than CBM, without making any defect state inside the energy gap [Figs.~4, 5(b), and 6(b)]. It is very intriguing that defect states associated with the hydrogen atoms are formed at about 7--8~eV below VBM and at about 3~eV above CBM, which are drawn in red in Figs.~4 and 5(b). Our calculations yield remarkably that the total number of states below VBM is the same before and after the hydrogen-atom adsorption despite the formation of the defect state below VBM. Thus one electron, which is from the valence bands and the hydrogen atom, moves to the conduction bands, resulting in electron-type doping to the host material. In this result, the accuracy of the numerical counting of the states and the electrons is very important, so we checked it very carefully \cite{ref46}, reaching the above conclusion. Moreover, since the defect states are quite below VBM or above CBM, the validity of this doping effect is not affected by the well-known band-gap underestimation of the DFT. In our results, one hydrogen atom produces one electron in the conduction bands, which is an ultimate efficiency much higher than the previous prediction of 4 \% for the case of hydrogen-atom adsorption to sulfur atoms~\cite{ref14,ref17}.

As mentioned above, defect states of the adsorbed hydrogen atom at the hollow site are not located inside the band gap, but they are quite below VBM or above CBM. We analyzed the wave functions of the defect states in MoTe$_2$ monolayer and found that the occupied defect state at 7--8~eV below VBM is a bonding state between the hydrogen 1$s$ state and the molybdenum 5$s$ and 4$d$ states [Fig.~\ref{fig5}(d)], while the unoccupied defect state at about 3~eV above CBM is an antibonding state [Fig.~\ref{fig5}(e)]. We also analyzed the CBM state in MoTe$_2$ after hydrogen-atom adsorption, finding that it is indeed an extended state consisting mainly of $d_{z^2}$ orbitals of Mo atoms [Fig.~\ref{fig5}(f)]. The charge density near the hydrogen atom [Fig.~\ref{fig5}(g)] is very close to the sum of the charge densities of an isolated hydrogen atom and the pristine MoTe$_2$, so we can hardly say that the hydrogen impurity is ionized. Thus we summarize the electron doping mechanism of the adsorbed hydrogen atom as follows. When a hydrogen atom is adsorbed to a $MX_2$ monolayer, it goes to the center of a hexagonal hollow site and forms bonding and antibonding states with surrounding three transition-metal atoms. Here the host material contributes effectively one state, out of the valence bands, to the bonding and antibonding states, as presented schematically in Fig.~\ref{fig5}(c). The bonding and antibonding states then would be filled with three electrons: two from the valence band and one from the hydrogen atom. However, the antibonding state is higher in energy than CBM, so the CBM state is filled with one electron instead of the antibonding state, leading to one electron doping per hydrogen atom (Fig.~\ref{fig7}). Furthermore, since the defect states are quite away from VBM and CBM, the scattering of electrons by the defect states will be very weak, supporting a high mobility of electrons in the conduction bands.

\begin{table*}[t] 
\caption
[Doping types and binding energies of MoTe$_2$ with defects.]
{Doping types and binding energies of MoTe$_2$ with defects. The doping types and binding energies of MoTe$_2$ monolayer, bilayer, and bulk are presented for the cases with tellurium vacancy, hydrogen atom (H) adsortption at a hollow site, hydrogen molecule (H$_2$) adsorption at a $M$-top site, and substitutional hydrogen atom (H), hydrogen molecule (H$_2$), and water molecule (H$_2$O) at a tellurium site. Here, our calculations are performed using a $5\times5$ supercell with one defect in it.}
\label{tab4}
\setlength{\tabcolsep}{9.4mm} 
\renewcommand{\arraystretch}{1.12}
\begin{tabular}{l c c c}
\hline\hline
Defect type  & Layer-thickness & Binding energy & Doping type \\
\hline
\multirow{3}*{Te vacancy} & Monolayer & & $h$ doping \\
& Bilayer & & $h$ doping \\
& Bulk & & $h$ doping \\
\hline
\multirow{2}*{H adsorption} & Monolayer & 2.614 eV/H & $e$ doping \\
& Bulk & 2.727 eV/H & $e$ doping \\
\hline
H$_2$ adsorption & Monolayer & 0.128 eV/H$_2$ & $h$ doping \\
\hline
\multirow{3}*{H substitution at Te site} & Monolayer & 3.701 eV/H & Nearly neutral \\
& Bilayer & 3.662 eV/H & Nearly neutral \\
& Bulk & 3.624 eV/H & Nearly neutral \\
\hline
\multirow{3}*{H$_2$ substitution at Te site} & Monolayer & 0.907 eV/H$_2$ & $h$ doping \\
& Bilayer & 0.918 eV/H$_2$ & $h$ doping \\
& Bulk & 1.076 eV/H$_2$ & $h$ doping \\
\hline
H$_2$O substitution at Te site & Monolayer & 0.488 eV/H$_2$O & $h$ doping\\
\hline\hline
\end{tabular}
\end{table*} 

The above mentioned doping mechanism by hydrogen-atom adsorption corresponds to the formation of donor levels above CBM as in bulk InN and ZnO \cite{ref30,ref33}. Thus, electrons in the donor levels simply drop to the conduction band so that the concentration of electron carriers in the conduction band is equal to the concentration of hydrogen atoms in the sample. Furthermore, the absence of the activation energy (or the presence of the negative activation energy) makes the number of electron carriers in the conduction band independent of temperature. Thus the temperature dependence of the electrical resistivity may behave like a metal rather than a semiconductor, i.e., the electrical resistivity may decrease as the temperature decreases because the electron-phonon scattering becomes less frequent at lower temperature.

\vspace{-2mm}
\subsection{Effects of H$_2$ adsorption and Te vacancy}
\vspace{-2mm}
For comparison, we also considered adsorption of the hydrogen molecule (H$_2$). We obtained atomic and electronic structures of $MX_2$ monolayers with a hydrogen molecule adsorbed at the hollow, $M$-top, and $X$-top sites. Among these, the $M$-top case is the lowest total-energy configuration and it produces hole doping in contrast to the electron doping by the hydrogen atom. Our calculation also shows that the total energy of the system decreases if two adsorbed hydrogen atoms meet each other and form a H$_2$ molecule. Thus, in order to maintain the electron doping effect of the hydrogen atom, adsorbed hydrogen atoms should stay away from one another, which is relevant because the hydrogen-atom diffusion is suppressed by the energy barriers described above.

Real samples of MoTe$_2$ are often $p$-type and it may be attributed to Te vacancies \cite{ref15}. Recently, it was reported that the presence of Te vacancies increases the binding energy of molecules such as O$_2$ by an order of magnitude \cite{ref10}. Thus the doping type of MoTe$_2$ with an adsorbed hydrogen atom at a Te vacancy needs to be clarified. We considered a hydrogen atom (H), a hydrogen molecule (H$_2$), and a water molecule (H$_2$O) placed at a Te vacancy in MoTe$_2$ monolayer. We obtained atomic and electronic structures of MoTe$_2$ with the defects as shown in Figs.~6 and 8. With the Te vacancy, the chemical potential \cite{ref47} becomes closer to VBM [Fig.~\ref{fig8}(c)], making $p$-type doping. In the case of the hydrogen atom at the Te vacancy, which is energetically preferred to a hydrogen molecule at the Te vacancy (Table~\ref{tab4}), the hydrogen atom compensates the hole doping generated by the Te vacancy, making MoTe$_2$ almost free of charge carriers [Fig.~\ref{fig8}(d)]. In the case of H$_2$ at the Te vacancy, the material remains hole-doped since the H$_2$ states are far away from the chemical potential and no charge transfer occurs from H$_2$ [Fig.~\ref{fig8}(e)]. In the case of H$_2$O at the Te vacancy, the material remains hole-doped [Fig.~\ref{fig8}(f)]. Band structures are shown in Fig.~\ref{fig6} and the binding energies and the doping types are summarized in Table~\ref{tab4}. Our results show that the hydrogen-atom doping is distinctively effective in compensating the $p$-type doping in MoTe$_2$ with Te vacancy, so $p$- to $n$-type conversion can be achieved by hydrogen-atom doping.

The hydrogen-atom doping can be attained in real samples, for example, by atomic layer deposition (ALD) of Al$_2$O$_3$, where hydrogen atoms can be released from the reaction of water molecules with aluminum \cite{ref28}. Thus, an initially $p$-doped channel in a single TMDC nanosheet can be selectively converted to $n$-doped region by patterned ALD, forming an electronic device in a single nanosheet.

\vspace{-2mm}
\section{Conclusion}
\vspace{-2mm}
In conclusion, we studied adsorbed hydrogen atoms in TMDC materials using first-principles calculations based on DFT. The spatial stability of the hydrogen atoms was demonstrated by calculating the binding energy and the diffusion barrier. Our result showed that the adsorbed hydrogen atom has the lowest energy at the center of the hexagonal hollow site right on the metal plane, doping one electron per hydrogen atom to the conduction bands. The diffusion barrier was found large in selenides and tellurides, stabilizing the dopants at their initial positions. We analyzed the bonding and antibonding states of the hydrogen atom and its neighboring metal atoms and presented the mechanism for its highly efficient $n$-type doping without any defect state inside the band gap. We also studied doping effects of hydrogen molecules and Te vacancies for comparison. Our results show that the hydrogen adatom is an excellent $n$-type dopant with negligible diffusion in TMDC nanosheets, opening a doping strategy for highly integrated nanometer-scale TMDC devices.

	\vspace{-2mm}
\begin{acknowledgments}
	\vspace{-2mm}
This work was supported by the NRF of Korea (Grant No. 2011-0018306). J.Y.L. and S.I. acknowledge financial support from NRF of Korea (Grants No. 2017R1A2A1A05001278 and No. 2017R1A5A1014862). Computational resources have been provided by KISTI Supercomputing Center (Project No. KSC-2016-C3-0006).
\end{acknowledgments}


\begin{thebibliography}{99}
\bibitem{ref1}
B. Radisavljevic, A. Radenovic, J. Brivio, V. Giacometti, and A. Kis, Nat. Nanotech. {\bf 6}, 147 (2011).
\bibitem{ref2}
H. Wang, L. Yu, Y.-H. Lee, Y. Shi, A. Hsu, M. L. Chin, L.-J. Li, M. Dubey, J. Kong, and T. Palacios, Nano Lett. {\bf 12}, 4674 (2012).
\bibitem{ref3}
Q. H. Wang, K. Kalantar-Zadeh, A. Kis, J. N. Coleman, and M. S. Strano, Nat. Nanotech. {\bf 7}, 699 (2012).
\bibitem{ref4}
D. H. Keum, S. Cho, J. H. Kim, D.-H. Choe, H.-J. Sung, M. Kan, H. Kang, J.-Y. Hwang, S. W. Kim, H. Yang, K. J. Chang, and Y. H. Lee, Nat. Phys. {\bf 11}, 482 (2015).
\bibitem{ref5}
H. S. Lee, S. S. Baik, K. Lee, S.-W. Min, P. J. Jeon, J. S. Kim, K. Choi, H. J. Choi, J. H. Kim, and S. Im, ACS Nano {\bf 9}, 8312 (2015).
\bibitem{ref6}
H.-P. Komsa, J. Kotakoski, S. Kurasch, O. Lehtinen, U. Kaiser, and A. V. Krasheninnikov, Phys. Rev. Lett. {\bf 109}, 035503 (2012).
\bibitem{ref7}
H. Fang, M. Tosun, G. Seol, T. C. Chang, K. Takei, J. Guo, and A. Javey, Nano Lett. {\bf 13}, 1991 (2013).
\bibitem{ref8}
Y.-C. Lin, D. O. Dumcenco, H.-P. Komsa, Y. Niimi, A. V. Krasheninnikov, Y.-S. Huang, and K. Suenaga, Adv. Mater. {\bf 26}, 2857 (2014).
\bibitem{ref9}
J. Suh, T.-E. Park, D.-Y. Lin, D. Fu, J. Park, H. J. Jung, Y. Chen, C. Ko, C. Jang, Y. Sun, R. Sinclair, J. Chang, S. Tongay, and J. Wu, Nano Lett. {\bf 14}, 6976 (2014).
\bibitem{ref10}
B. Chen, H. Sahin, A. Suslu, L. Ding, M. I. Bertoni, F. M. Peeters, and S. Tongay, ACS Nano {\bf 9}, 5326 (2015).
\bibitem{ref11}
C.-H. Lee, G.-H. Lee, A. M. van der Zande, W. Chen, Y. Li, M. Han, X. Cui, G. Arefe, C. Nuckolls, T. F. Heinz, J. Guo, J. Hone, and P. Kim, Nat. Nanotech. {\bf 9}, 676 (2014).
\bibitem{ref12}
A. Pezeshki, S. H. H. Shokouh, P. J. Jeon, I. Shackery, J. S. Kim, I.-K. Oh, S. C. Jun, H. Kim, and S. Im, ACS Nano {\bf 10}, 1118 (2016).
\bibitem{ref13}
Z. Lin, B. R. Carvalho, E. Kahn, R. Lv, R. Rao, H. Terrones, M. A. Pimenta, and M. Terrones, 2D Mater. {\bf 3}, 022002 (2016).
\bibitem{ref14}
J. G. He, K. C. Wu, R. J. Sa, Q. H. Li, and Y. Q. Wei, Appl. Phys. Lett. {\bf 96}, 082504 (2010).
\bibitem{ref15}
Y. Ma, Y. Dai, M. Guo, C. Niu, J. Lu, and B. Huang, Phys. Chem. Chem. Phys. {\bf 13}, 15546 (2011).
\bibitem{ref16}
D. Voiry, R. Fullon, J. Yang, C. C. C. Silva, R. Kappera, I. Bozkurt, D. Kaplan, M. J. Lagos, P. E. Batson, G. Gupta, A. D. Mohite, L. Dong, D. Er, V. B. Shenoy, T. Asefa, and M. Chhowalla, Nat. Mater. {\bf 15}, 1003 (2016).
\bibitem{ref17}
Y. Kim, Y. I. Jhon, J. Park, C. Kim, S. Lee, and Y. M. Jhon, Sci. Rep. {\bf 6}, 21405 (2016).
\bibitem{ref18}
L. K. Li, Y. J. Yu, G. J. Ye, Q. Q. Ge, X. D. Ou, H. Wu, D. L. Feng, X. H. Chen, and Y. B. Zhang, Nat. Nanotech. {\bf 9}, 372 (2014).
\bibitem{ref19}
J. Kim, S. S. Baik, S. H. Ryu, Y. Sohn, S. Park, B.-G. Park, J. Denlinger, Y. Yi, H. J. Choi, and K. S. Kim, Science {\bf 349}, 723 (2015).
\bibitem{ref20}
S. S. Baik, K. S. Kim, Y. Yi, and H. J. Choi, Nano Lett. {\bf 15}, 7788 (2015).
\bibitem{ref21}
H. Doh and H. J. Choi, 2D Mater. {\bf 4}, 025071 (2017).
\bibitem{ref22}
Y. Yan, B. Xia, Z. Xu, and X. Wang, ACS Catal. {\bf 4}, 1693 (2014).
\bibitem{ref23}
J. Qi, Y.-W. Lan, A. Z. Stieg, J.-H. Chen, Y.-L. Zhong, L.-J. Li, C.-D. Chen, Y. Zhang, and K. L. Wang, Nat. Commun. {\bf 6}, 7430 (2015).
\bibitem{ref24}
M. M. Aly\"or\"uk, Y. Aierken, D. \c Cakir, F. M. Peeters, and C. Sevik, J. Phys. Chem. C {\bf 119}, 23231 (2015).
\bibitem{ref25}
Z. Yin, H. Li, H. Li, L. Jiang, Y. Shi, Y. Sun, G. Lu, Q. Zhang, X. Chen, and H. Zhang, ACS Nano {\bf 6}, 74 (2012).
\bibitem{ref26}
O. Lopez-Sanchez, D. Lembke, M. Kayci, A. Radenovic, and A. Kis, Nat. Nanotech. {\bf 8}, 497 (2013).
\bibitem{ref27}
X. Xu, W. Yao, D. Xiao, and T. F. Heinz, Nat. Phys. {\bf 10}, 343 (2014).
\bibitem{ref28}
J. Y. Lim, A. Pezeshki, S. Oh, J. S. Kim, Y. T. Lee, S. Yu, D. K. Hwang, G.-H. Lee, H. J. Choi, and S. Im, Adv. Mater {\bf 29}, 1701798 (2017).
\bibitem{ref29}
R. C. Newman, Rep. Prog. Phys. {\bf 45}, 1163 (1982).
\bibitem{ref30}
C. G. van de Walle and J. Neugebauer, Annu. Rev. Mater. Res. {\bf 36}, 179 (2006).
\bibitem{ref31}
U. M. Gosele, Annu. Rev. Mater. Sci. {\bf 18}, 257 (1988).
\bibitem{ref32}
S. J. Pearton, J. W. Corbett, and T. S. Shi, Appl. Phys. A {\bf 43}, 153 (1987).
\bibitem{ref33}
C. G. Van de Walle, Phys. Rev. Lett. {\bf 85}, 1012 (2000).
\bibitem{ref34}
K. J. Chang and D. J. Chadi, Phys. Rev. B {\bf 40}, 11644 (1989).
\bibitem{ref35}
J. P. Perdew, K. Burke, and M. Ernzerhof, Phys. Rev. Lett. {\bf 77}, 3865 (1996).
\bibitem{ref36}
N. Troullier and J. L. Martins, Phys. Rev. B {\bf 43}, 1993 (1991).
\bibitem{ref37}
J. M. Soler, E. Artacho, J. D. Gale, A. García, J. Junquera, P. Ordej\'on, and D. S\'anchez-Protal, J. Phys.: Condens. Matter {\bf 14}, 2745 (2002).
\bibitem{ref38}
L. Kleinman, Phys. Rev. B {\bf 21}, 2630 (1980).
\bibitem{ref39}
G. Theurich and N. A. Hill, Phys. Rev. B {\bf 64}, 073106 (2001).
\bibitem{ref40}
L. Kleinman and D. M. Bylander, Phys. Rev. Lett. {\bf 48}, 1425 (1982).
\bibitem{ref41}
C.-Y. Moon, J. Han, H. Lee, and H. J. Choi, Phys. Rev. B {\bf 84}, 195425 (2011).
\bibitem{ref42}
S. Oh and H. J. Choi, Sci. Rep. {\bf 7}, 2024 (2017).
\bibitem{ref43}
G. Kresse and J. Hafner, Phys. Rev. B {\bf 47}, 558 (1993).
\bibitem{ref44}
G. Kresse and J. Furthm\"uller, Phys. Rev. B {\bf 54}, 11169 (1996).
\bibitem{ref45}
By using VASP, we obtained the binding energy of the hydrogen atom adsorbed at various sites in monolayer MoS$_2$ and MoTe$_2$. Our VASP results also show the hydrogen atom adsorbed at the hollow site on the metal plane is the lowest energy configuration, confirming our SIESTA results.
\bibitem{ref46}
We used the standard method to count the number of electrons. The Fermi-Dirac occupation number of each state is summed for the band index $n$ and the wave vector {\bf k}, divided by the number of {\bf k} points.
\bibitem{ref47}
When a system has a finite energy difference ($\Delta E$) between highest occupied states ($E_1$) and lowest unoccupied states ($E_2$), the chemical potential determined by the Fermi-Dirac distribution can deviate by the order of $k_BT$ from ($E_1+E_2$)/2. Here $k_B$ is Boltzmann constant and $T$ is the temperature of the system. In our present work, we considered low-temperature limit ($T\ll\Delta E/k_B$) so that the chemical potential approaches ($E_1+E_2$)/2, which is equal to the center of the energy gap in pristine case, the average of VBM and the lowest acceptor level in the $p$-doped case, and the average of the highest donor level and CBM in the $n$-doped case.
\end{thebibliography}
\end{document}